\documentclass[preprint]{aastex62}
\usepackage{graphicx}
\usepackage{amsmath}
\graphicspath{{/Users/nick/mypapers/2019_aj_irtf/images/}}

\begin{document}
	
	\title{\Large{Detection of Propadiene on Titan}}
	
	\shorttitle{Detection of Propadiene}
	\shortauthors{Lombardo et al.}
	
	\correspondingauthor{Nicholas A Lombardo}
	\email{nicholas.lombardo@yale.edu}
	
	\author{Nicholas A Lombardo \thanks{Corresponding author now at Yale University, Department of Geology and Geophysics}}
	\affiliation{Center for Space Science and Technology, University of Maryland, Baltimore County, 1000 Hilltop Circle, Baltimore, MD, USA}
	\affiliation{Goddard Space Flight Center, 8800 Greenbelt Road, Greenbelt, MD, 20770, USA}
	\affiliation{Department of Geology and Geophysics, Yale University, New Haven, CT, 06511, USA}
	
	\author{Conor A Nixon}
	\affiliation{NASA Goddard Space Flight Center, 8800 Greenbelt Road, Greenbelt, MD, 20770, USA}
	
	\author{Thomas K Greathouse}
	\affiliation{Department of Space Science, Southwest Research Institute, 6220 Culebra Road, San Antonio, TX, 78228, USA}
	
	\author{Bruno B\'ezard}
	\affiliation{LESIA, Observatoire de Paris, Universit\'e PSL, CNRS, Sorbonne Universit\'e, Universit\'e de Paris, 5 place Jules Janssen, 92195 Meudon, France}
	
	\author{Antoine Jolly}
	\affiliation{LISA, UMR 7583, Universit\"e Paris Est Creteil, Universit\'e de Paris, IPSL, Creteil, France}
	
	\author{Sandrine Vinatier}
	\affiliation{LESIA, Observatoire de Paris, Universit\'e PSL, CNRS, Sorbonne Universit\'e, Universit\'e de Paris, 5 place Jules Janssen, 92195 Meudon, France}
	
	\author{Nicholas A Teanby}
	\affiliation{School of Earth Sciences, University of Bristol, UK}
	
	\author{Matthew J Richter}
	\affiliation{University of California, Davis, Davis, CA, USA}
	
	\author{Patrick J G Irwin}
	\affiliation{Atmospheric, Oceanic and Planetary Physics, Clarendon Laboratory, University of Oxford, Parks Road, Oxford OX1 3PU, UK}
	
	\author{Athena Coustenis}
	\affiliation{LESIA, Observatoire de Paris, Universit\'e PSL, CNRS, Sorbonne Universit\'e, Universit\'e de Paris, 5 place Jules Janssen, 92195 Meudon, France}
	
	\author{F Michael Flasar}
	\affiliation{NASA Goddard Space Flight Center, 8800 Greenbelt Road, Greenbelt, MD, 20770, USA}
	
	\begin{abstract}
	The atmosphere of Titan, the largest moon of Saturn, is rich in organic molecules, and it has been suggested that the moon may serve as an analog for the pre-biotic Earth due to its highly reducing chemistry and existence of global hazes.  Photochemical models of Titan have predicted the presence of propadiene (historically referred to as allene), CH$_{2}$CCH$_{2}$, an isomer of the well-measured propyne (also called methylacetylene) CH$_{3}$CCH, but its detection has remained elusive due to insufficient spectroscopic knowledge of the molecule - which has recently been remedied with an updated spectral line list.  Here we present the first unambiguous detection of the molecule in any astronomical object, observed with the Texas Echelle Cross Echelle Spectrograph (TEXES) on the NASA Infrared Telescope Facility (IRTF) in July 2017.  We model its emission line near 12 $\mu$m and measure a volume mixing ratio (VMR) of (6.9 $\pm$ 0.8) $\times$10$^{-10}$ at 175 km, assuming a vertically increasing abundance profile as predicted in photochemical models.  Cassini measurements of propyne made during April 2017 indicate that the abundance ratio of propyne to propadiene is 8.2$\pm$1.1 at the same altitude. This initial measurement of the molecule in Titan's stratosphere paves the way towards constraining the amount of atomic hydrogen available on Titan, as well as future mapping of propadiene on Titan from 8 meter and larger ground based observatories, and future detection on other planetary bodies.
	\end{abstract}

	
	\section{Introduction}
	
\begin{figure*}
		\includegraphics[width = \textwidth]{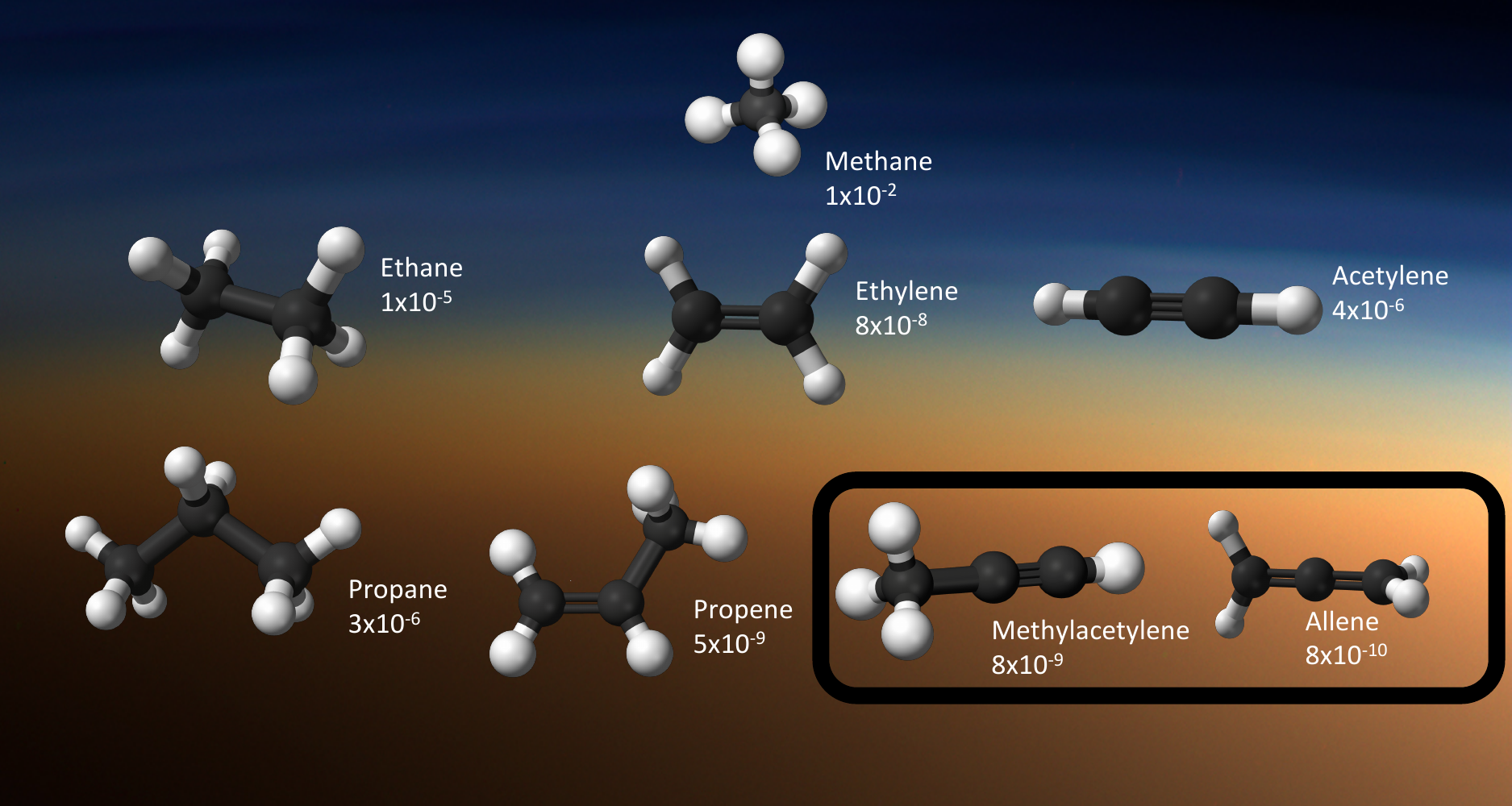}
		\caption{Neutral C1, C2, and C3 hydrocarbons of Titan's atmosphere and their approximate volume mixing ratios at 200 km derived from \cite{vinatier:2015}.  Heavier species are not shown.  Propyne and propadiene now constitute the first hydrocarbon isomer pair detected on Titan. \label{fig:molecules}}
	\end{figure*}

	Nearly 98\% of Titan's stratosphere by volume is molecular nitrogen.  The secondary species, methane, comprises between 1.1 and 1.4 \% of the stratosphere \cite{lellouch:14}.  The fragmentation of these molecules by solar UV photons and electrons excited by Saturn's magnetosphere, and successive reactions between the resultant radicals and ions, gives rise to species of the form C$_{x}$H$_{y}$ and C$_{x}$H$_{y}$N$_{z}$ - some of which are shown in Figure \ref{fig:molecules}.  Knowledge of the production and depletion pathways for Titan's hydrocarbons and nitrogen-bearing species is critical for understanding the production of Titan's hazes.
	
	The Cassini spacecraft explored the Saturn system from 2004 through 2017, during which it made 127 targeted flybys of Titan.  On-board the spacecraft were a suite of remote sensing instruments, including the Composite Infrared Spectrometer (CIRS; \cite{jennings:17}).  CIRS was capable of performing spatially resolved observations of Titan at spectral resolutions from 0.5 cm$^{-1}$ to 15.5 cm$^{-1}$ ($\lambda$/$\Delta\lambda \approx$ 40 - 3000), and observations from the instrument helped to elucidate the complex chemistry and dynamics of Titan's atmosphere including the first detection of propene, C$_{3}$H$_{6}$ \citep{nixon:propene, vinatier:2015, coustenis:18, teanby:19, lombardo:2019b}.  
	
	With the successful end to the Cassini mission in 2017, ground-based observations have become the only vehicle by which continuing studies of Titan's atmosphere can be made.  Recent studies of Titan using the capabilities of the Atacama Large sub-Millimeter Array (ALMA) have resulted in detections of new species, including C$_{2}$H$_{5}$CN \citep{cordiner:15} and C$_{2}$H$_{3}$CN \citep{palmer:17}, new isotopologues \citep{serigano:16, molter:16, thelen:ch3d} and the ability to continue studying the spatial and temporal variations of Titan's atmosphere \citep{lai:17, cordiner:18, cordiner:19, teanby:18, thelen:19}.  Many molecular species may be studied using sub-mm observations, which are sensitive to the rotational transitions of small polar molecules. Symmetric molecules (including propadiene), however, lacking an intrinsic dipole, may only be studied in the infrared via their vibrational transitions.
	
	TEXES \citep{lacy:texes} is a ground-based mid-infrared spectrograph capable of achieving a resolving power of $\lambda$/$\Delta\lambda \approx$ 50,000 - 100,000 - about 50 times higher resolution than CIRS - from 4.5 - 20 $\mu$m and 22 - 25 $\mu$m.  This increased spectral resolution allows TEXES to disentangle emission from closely spaced molecular transitions, enabling a more precise study of some of Titan's less abundant species, such as propene and propane.  TEXES has been used to observe Titan in the past, including the detection of Titan's mesosphere \citep{griffith:05}, and the first spectrally resolved measurement of propane on Titan \citep{roe:03}. 
	
	Propadiene, an isomer of propyne, has been predicted to be present in Titan's atmosphere for several decades, with a predicted abundance about one-tenth of propyne's \citep{ strobel:74, yung:1984, coustenis:2003}.  Though several searches have been performed in attempts to identify the molecule on Titan, none have definitively detected it \citep{coustenis:2003, nixon:allene, lombardo:2019}.  A tentative detection was reported \citep{roe:2011}, however it was discovered that the spectral line list used to identify the possible propadiene emission was not suitable for the high resolution TEXES observations used in the search.  In this letter, we make use of a newly created line list of propadiene \citep{lombardo:2019} to make the first unambiguous detection and measurement of the molecule on Titan, and in an extraterrestrial environment.
	
	\section{Observations}
	Observations of Titan were performed with TEXES mounted on the IRTF on 11 July 2017.  The 3-m primary mirror on IRTF gives a diffraction-limited angular resolution on the sky of about 0.7 arcsec in the mid-infrared, however this is often degraded by non-ideal seeing during the observations.  In July 2017, Titan's angular diameter was 0.75 arcsec, limiting our measurements to disc-averaged values.  We observed Titan using the high spectral resolution mode available on TEXES, which gave a spectral resolution of about 0.009 cm$^{-1}$ ($\lambda / \Delta \lambda \approx$ 80,000).  The 838.182 - 845.415 cm$^{-1}$ spectrum presented in this letter was acquired with a 410 second integration time broken up over many shorter 4.5 second exposures.
	
	Observations were performed by nodding Titan by 2.5 arcsec along the 1.4 x 7 arcsec slit on TEXES.  Flux calibration was performed following \cite{lacy:texes}, with a room temperature radiometric blackbody card measured before each observation file.  The observations were calibrated, geometrical distortions removed, and coadded using the TEXES pipeline software package (see the final extracted spectrum in Figure \ref{fig:12umfit}).  The Titan spectrum was divided by a nodded observation of the asteroid Juno to remove the effects of telluric absorption on the Titan spectrum.
	
	To contemporaneously measure the abundance of propyne on Titan, which is not easily seen from Earth due to poor telluric transmission near the molecule's $\nu_{9}$ band centered at 633 cm$^{-1}$, we also employ CIRS spectra observed during the T126 Cassini flyby of Titan which occurred in April 2017.  Both limb and nadir spectra were acquired at a distance of 2.6 - 4.8 $\times$10$^{5}$ km from Titan.  The two mid-infrared detector arrays were pointed at Titan and slewed across the surface, enabling global coverage comparable to disk-averaged ground based observations.  The spectrum we model was calculated from a total of 57448 individual spectra at a spectral resolution of 3 cm$^{-1}$.
	
	\section{Modeling}
	Spectral modeling of the observations was performed with the Nonlinear optimal Estimator for MultispEctral analySIS (NEMESIS, \cite{irwin:nemesis}), which has been used to model planetary atmospheres extensively \citep{teanby:13, irwin:18, lombardo:2019, lombardo:2019b, teanby:19}.  NEMESIS uses an iterative inversion scheme that calculates the `best-fit' of a synthetic to observed spectrum.  This is achieved by varying \textit{a priori} profiles of modeled quantities (for example temperature or molecular abundance) derived from previous observations.  The final measured values are those that simultaneously minimize the difference between the synthetic and observed spectrum, and the deviation of the measured profiles from the \textit{a priori} values.  The extent of the atmosphere to which our data is most sensitive to can be approximated by the Full-Width Half-Max of the contribution function.  The contribution function for each molecule at a specific wavenumber can be defined as the change of model radiance at that wavenumber with respect to the amount of the molecule present in the model - dI/dq, where I is the radiance and q is the abundance, shown in Figure \ref{fig:c3plot}C.
	
	The observations we model are sensitive to a limited region in Titan's atmosphere, about 100 km - 230 km, so we allow the \textit{a priori} profiles to vary only by uniform scaling factors.
	
	The correlated-k approximation was used to calculate the spectra during the retrieval.  Spectral line data for CH$_{4}$ and C$_{2}$H$_{6}$ are from the HITRAN 2016 database.  Spectral line data for C$_{3}$H$_{8}$ are from pseudo-line lists reported in \cite{sung:propane}.  The updated spectral line list for propadiene is described in \cite{lombardo:2019}.
	
\begin{figure}
		\includegraphics[width=\textwidth]{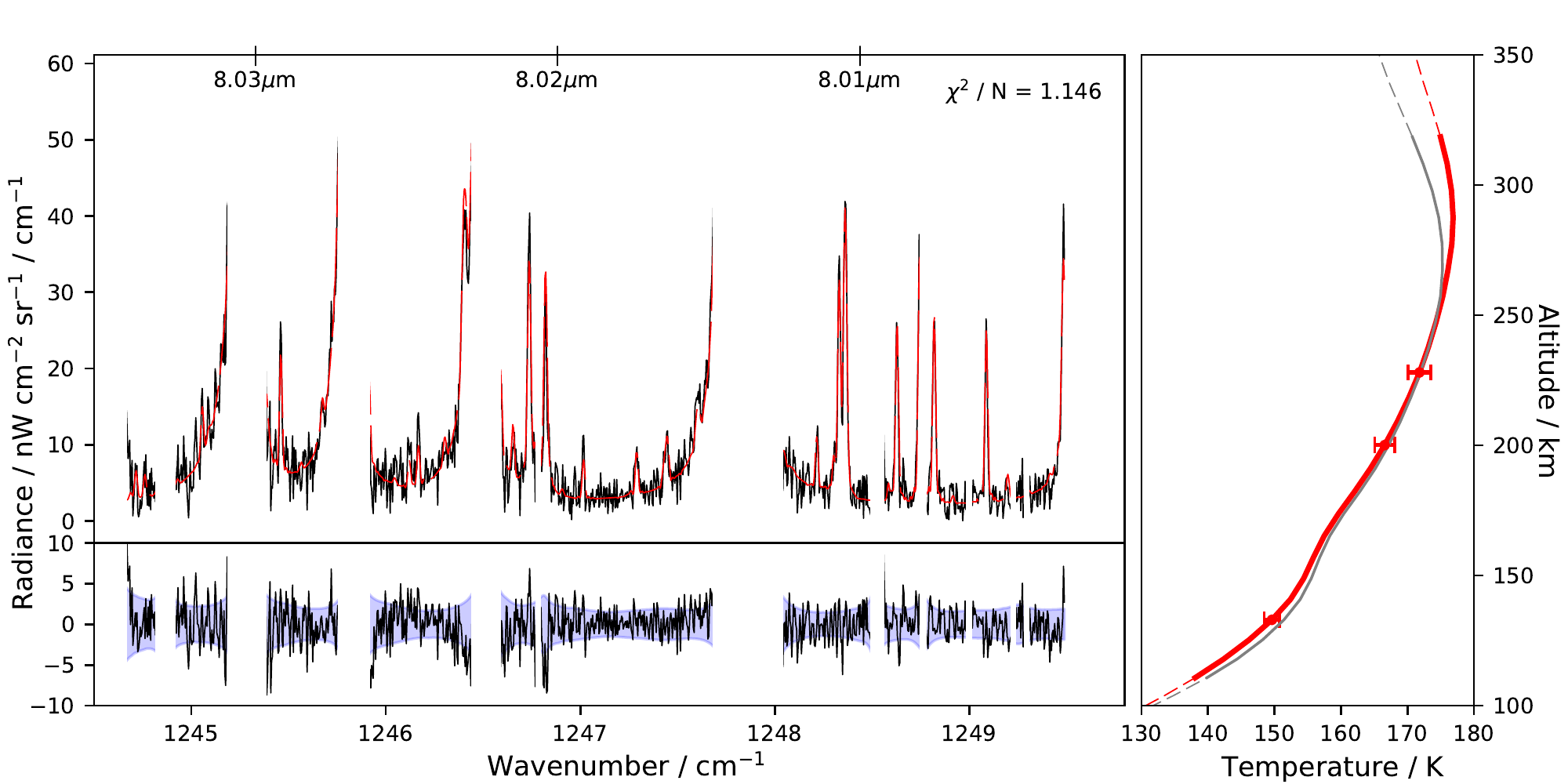}
		\caption{The fit of the synthetic (red) to the observed (black) TEXES spectrum used to extract the thermal profile of Titan at the time of observation.  Corresponding wavelengths in microns are shown for convenience on the upper horizontal axis.  The spectral noise on the data is shown in blue.  The gray profile is the \textit{a priori} profile derived from the T126 flyby. \label{fig:temp}}
	\end{figure}

	To determine Titan's thermal structure used in the 12-$\mu$m observations, we also observed the spectral region 1244.5 - 1249.5 cm$^{-1}$ with TEXES, which includes emission from the $\nu_{4}$ band of methane.  The VMR of methane in Titan's stratosphere is well constrained, though it has been shown to vary with latitude \cite{lellouch:14}, allowing us to retrieve a vertical temperature profile with NEMESIS.  We utilize an \textit{a priori} temperature profile derived from CIRS observations during the T126 flyby of Titan in April 2017, determined using the same method as in \cite{achterberg:2014}.  The temperature profile we retrieve is very similar to the \textit{a priori} profile derived from CIRS data.  The retrieved thermal profile and fit of the synthetic to the observed spectrum is shown in Figure \ref{fig:temp}.
	
	The \textit{a priori} profile for C$_{2}$H$_{6}$ was set to (1.0$\pm$0.5)$\times$10$^{-5}$ throughout the stratosphere, condensing near 64 km and following its saturation vapor pressure curve below.  The retrieved scaling factor was 1.1$\pm$0.02, so our measurement is therefore (1.1$\pm$0.02)$\times$10$^{-5}$.  C$_{3}$H$_{8}$ was set to 1.0$\times$10$^{-6}$ in the stratosphere, and condensed at the same altitude as ethane.	We retrieve a scaling factor of 1.03$\pm$0.47. For propadiene, which has not been measured before, we test two profiles - a vertically uniform profile of 1$\times$10$^{-9}$ and a vertically increasing profile derived from \cite{vuitton:19}.  The scaling factor retrieved for the uniform profile is 0.84$\pm$0.10, and the scaling factor for the vertically increasing profile is 2.22$\pm$0.26.
	
	To contemporaneously measure the more abundant propyne isomer, we employ observations of the 633 cm$^{-1}$ $\nu_{9}$ band of the molecule observed by CIRS during T126 \citep{nixon:19} .  Observations of Titan by CIRS were made at a spectral resolution of 3 cm$^{-1}$, significantly less than the 0.01 cm$^{-1}$ resolution afforded by TEXES.  Though CIRS was capable of achieving a resolution of 0.5 cm$^{-1}$, to ensure global coverage of Titan during a flyby a faster but lower resolution instrument setting was used. The lower resolution nadir CIRS spectra are sensitive to the region between 80 km and 175 km, lower than, but overlapping, the range of propadiene sensitivity from approximately 100 km to 230 km.  The CH$_{3}$CCH \textit{a priori} profile was set to the prediction by \cite{vuitton:19}.  We retrieved a scale factor of 1.06$\pm$0.08. 
	
	\section{Results}
\begin{figure*}
	\includegraphics[width = \textwidth]{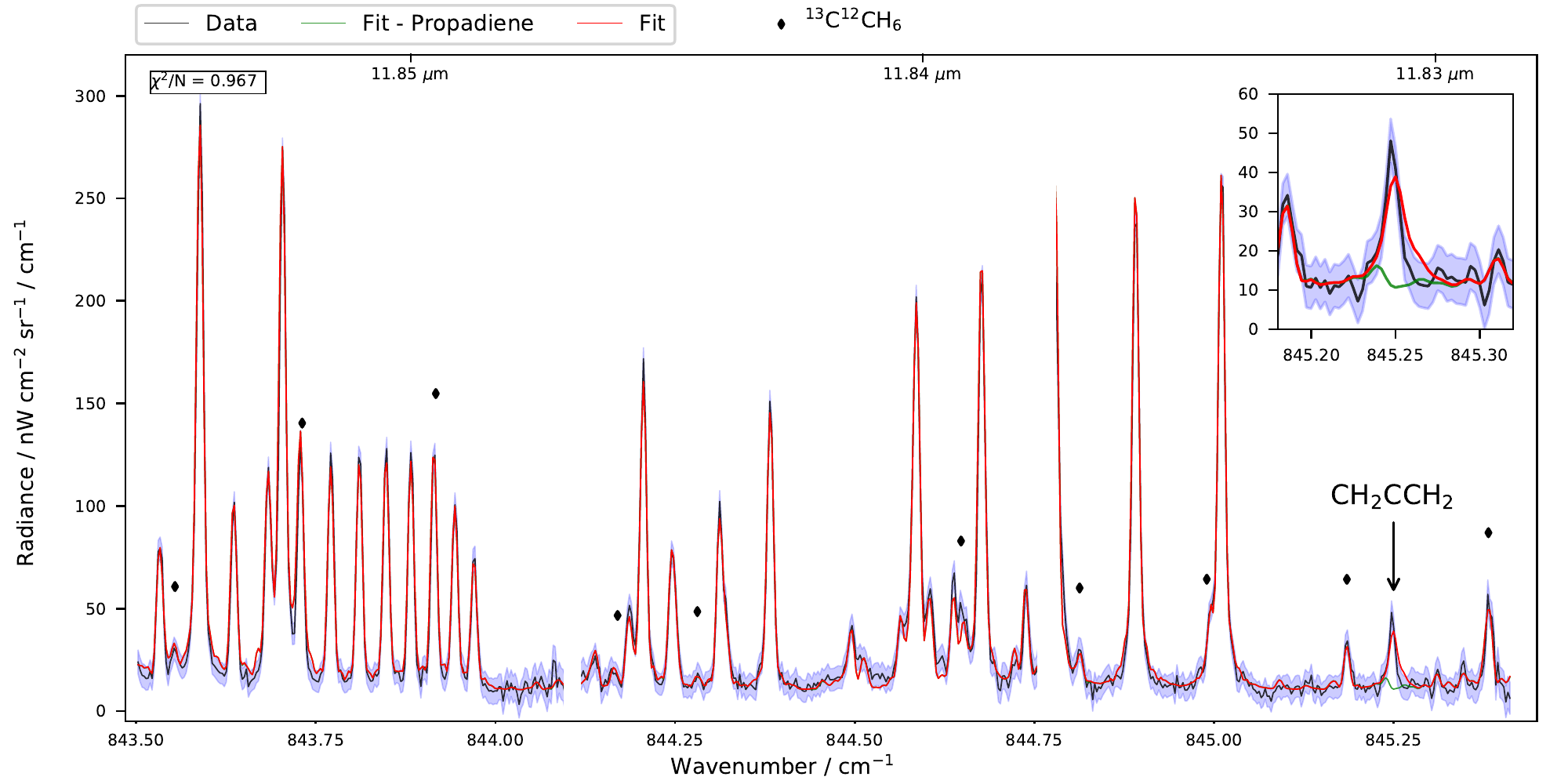}
	\caption{Model fit of the synthetic (red) to observed (black) TEXES spectrum near 12 $\mu$m, including the 1-$\sigma$ error envelope in blue.  Corresponding wavelengths in microns are shown on the upper horizontal axis.  The green line is the fit with emission from propadiene not included in the model  to clearly show the emission from propadiene.  The brightest non-labelled emission features are from C$_{2}$H$_{6}$.  The inset in the top right shows an enlarged view of the propadiene emission.  Prominent emission from $^{13}$C$^{12}$CH$_{6}$ are denoted by diamonds.  C$_{3}$H$_{8}$ is also included in the model, however it does not show any noticeable emission features in this region.  \label{fig:12umfit}}
	\end{figure*}

\begin{figure*}
	\includegraphics[width = \textwidth]{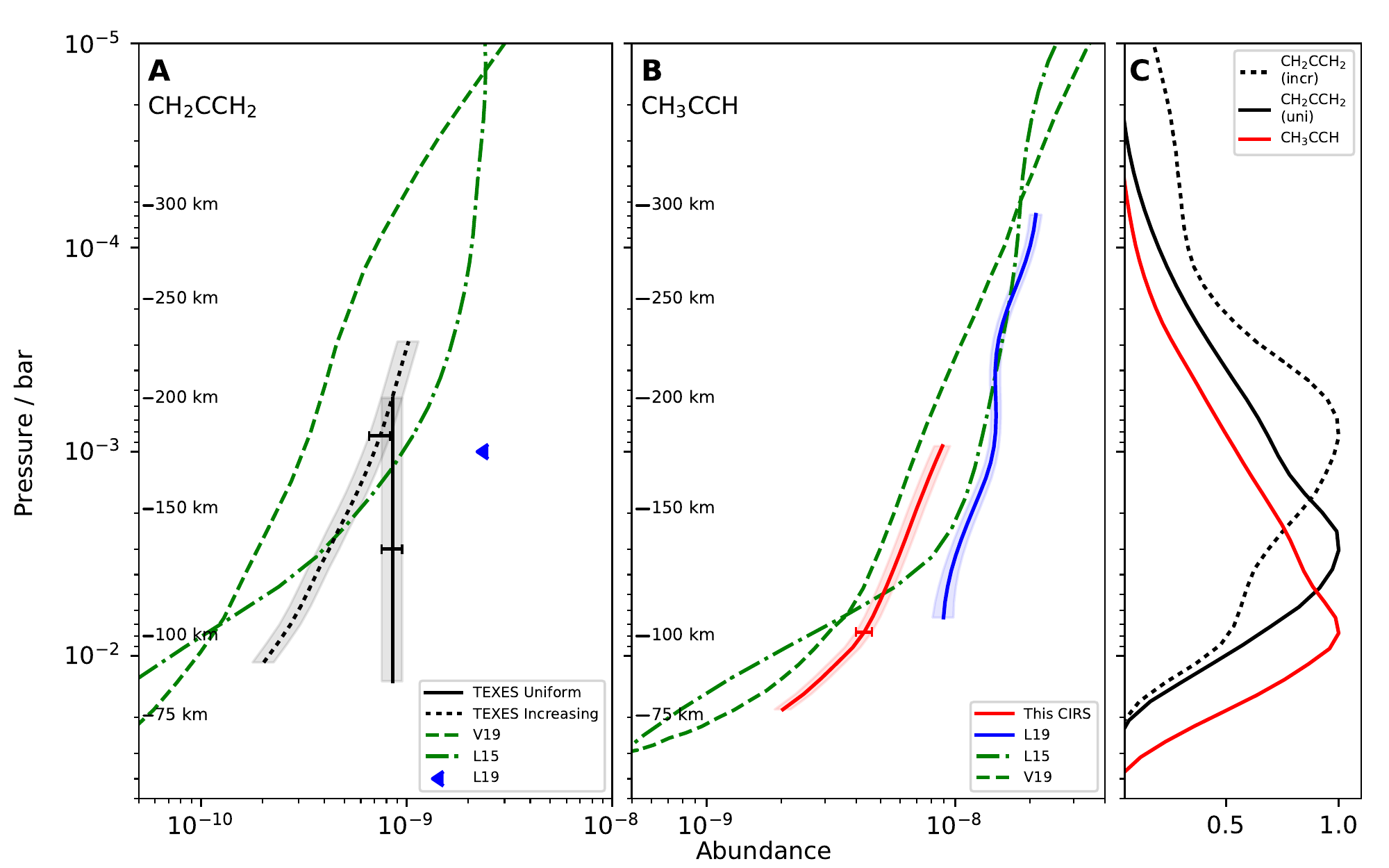}
	\caption{Retrieved abundance profiles of propadiene (left) and propyne (right) studied in this letter. Error bars are indicated at the peak of the contribution function.  If we assume that propadiene has a positive vertical gradient, then our measurement is sensitive to a higher altitude than if we assume a vertically uniform profile, shown in Panel C.  Photochemical model predictions from \cite{vuitton:19} and \cite{li:2015} are labeled V19 and L15, respectively.  L19 indicates measurements from \cite{lombardo:2019} described in the text. \label{fig:c3plot}}
	\end{figure*}

	The fit to the 12-$\mu$m TEXES spectrum is shown in Figure \ref{fig:12umfit}.  Gaps in spectral coverage indicate wavenumber regions where regions where the spectral orders are slightly larger than the TEXES detector and thus leave small gaps in spectral coverage.  These regions are not included in the model.  The green line indicates the model spectrum without propadiene - clearly showing the contribution from the molecule in the observations.  Due to the predominantly nadir view of these disc-averaged measurements and limited number of propadiene lines in the observation, we are unable to extract vertical information about the abundance.  For each of the molecules, our observations are most sensitive between 10$^{-2}$ and 10$^{-4}$ bar, indicated by the contribution functions in Figure \ref{fig:c3plot}C.  We test two different \textit{a priori} profiles for propadiene - a vertically constant profile at 1 ppbv and a vertical increasing profile derived from \cite{vuitton:19}.  We are able to acceptably model the observations using either of these profile shapes, with no significant difference in the quality of the spectral fit - indicating that we are unable to extract vertical information of the abundance of propadiene from these observations. The retrieved profiles for each the vertically constant and increasing profiles are shown in Figure \ref{fig:c3plot}A.

\begin{figure}
	\includegraphics[width = 0.8\columnwidth]{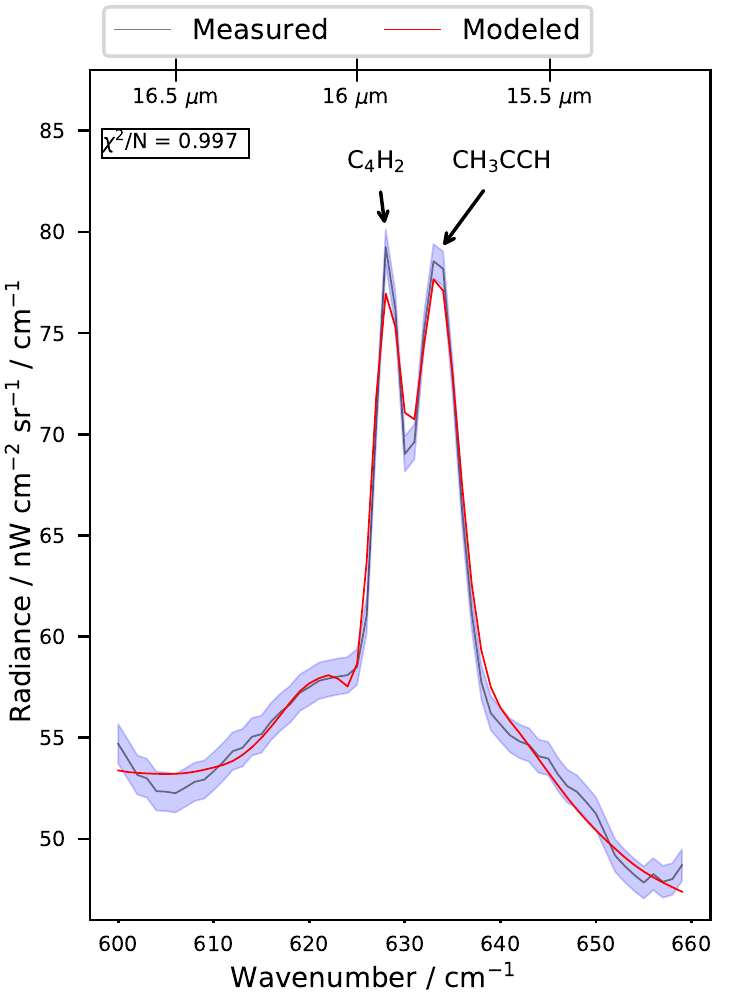}
	\caption{Model fit of the synthetic (red) to observed (black) CIRS spectrum used to retrieve the propyne abundance.  Corresponding wavelengths in microns are labeled for convenience on the upper horizontal axis.  \label{fig:cirsfit}}
	\end{figure}

	Figure \ref{fig:cirsfit} shows the fit of the T126 Cassini CIRS spectrum.  The spectral resolution is 3 cm$^{-1}$, oversampled to 1 cm$^{-1}$.  The choice to observe Titan at a lower spectral resolution decreases the time needed to perform the observation during the flyby.  At this resolution, emission from C$_{4}$H$_{2}$ and CH$_{3}$CCH are unresolved, and as both molecules have generally vertically varying abundance profiles, accurately fitting the spectrum while retrieving mixing ratios can be difficult.  As shown in Figure  \ref{fig:c3plot}B and C, the low spectral resolution of the observations only allows us to sample Titan's lower stratosphere, mainly between 90 km - 175 km.  propyne has been shown in previous works to have a strongly increasing abundance with altitude, hence we chose to use an increasing \textit{a priori} profile \cite{vuitton:19}.  Our measurements are consistent with predictions from \cite{li:2015} and \cite{vuitton:19}.  The discrepancy between the higher end of our measurements and the profiles reported in \cite{lombardo:2019} can be attributed to the disk-averaged nature of these measurements versus the long-term latitude binning \cite{lombardo:2019} used.
	
	\section{Discussion}
	We have provided the first definite detection and measurement of propadiene in Titan's stratosphere, measuring the molecule at an abundance of (6.9 $\pm$ 0.8) $\times$10$^{-10}$ at 175 km assuming a vertically increasing profile - consistent with previously reported upper-limits of 2.5$\times$10$^{-9}$ \cite{coustenis:2003, lombardo:2019}.  We also test a vertically uniform profile, which yields an abundance of 8.5$\pm$1.0 $\times$10$^{-10}$ and achieves a contribution function maximum at a deeper altitude, 135 km compared to 190 km.  The vertically increasing measurement falls between two predictions from photochemical models \cite{li:2015, vuitton:19} which both predict a vertically increasing profile. The ratio between the abundances of propadiene and propyne at 150 km is calculated to be 8.2 $\pm$ 1.1 if the vertically increasing profile is assumed.
	
	Both isomers of C$_{3}$H$_{4}$ have been included in several photochemical models \cite{li:2015, vuitton:19, hebrard:2013, dobrijevic:2016}.  Between 400 km and 800 km, CH$_{2}$CCH$_{2}$ is primarily produced by the addition of hydrogen to propargyl radicals (C$_{3}$H$_{3}$)
	\begin{equation}
	\label{}
	H + C_{3}H_{3} \rightarrow CH_{2}CCH_{2}
	\end{equation}
	This reaction persists to below 200 km, where the molecule can be produced additionally by the photodissociation of propene, 
	
	\begin{equation}
	\label{}
	C_{3}H_{6} + h\nu \rightarrow CH_{2}CCH_{2} + 2H
	\end{equation}
	Near 100 km, the production of propadiene via Equation 1 achieves its greatest value, which is also greater than the highest production rate achieved by Equation 2, albeit for a very narrow altitude range.
	
	The primary loss channels for CH$_{2}$CCH$_{2}$ between 100 km and 900 km are
	
	\begin{equation}
	\label{}
	CH_{2}CCH_{2} + h\nu \rightarrow C_{3}H_{3} + H
	\end{equation}
	\begin{equation}
	\label{}
	H + CH_{2}CCH_{2}  \rightarrow CH_{3}CCH + H
	\end{equation}
	with Equation 3 dominating from 100 km to 500 km and Eq. 4 above 500 km, though both reactions persists throughout this entire region \citep{vuitton:19}.

  	The hydrogen exchange reaction (Equation 4) is important in understanding the role of atomic hydrogen in Titan's stratosphere.  \cite{li:2015} discuss the potential use of the ratio between the abundance of propyne and propadiene as a probe of the concentration of atomic hydrogen available on Titan.  Equation 4 is highly exothermic, about 1.1 kcal/mol, which is a potential explanation for why the molar mixing ratio of propadiene is about an order of magnitude less than the mixing ratio of propyne in the stratosphere.  Below 200 km, a variation of Equation 4 exists such that instead of isomerizing propadiene, hydrogen will efficiently react with the molecule to form C$_{3}$H$_{5}$.  The production of C$_{3}$H$_{5}$ by hydrogen addition is also the final sink for propyne in the lower stratosphere.

	Titan's atmosphere includes both saturated (alkanes) and unsaturated (alkenes and alkynes) hydrocarbons.  The existence of unsaturated hydrocarbons on Titan requires a sink for atomic hydrogen, as atomic hydrogen will readily saturate unsaturated species.  A potential sink for H is descending aerosol particles.  As these particles descend through the atmosphere from their source hundreds of kilometers above Titan's surface, hydrogen may bond with the aerosol surfaces, an idea that was explored in \cite{sekine:08}.
	
	The measurements we perform in this paper are sensitive to the region between 100 km and 230 km above Titan's surface.  According to Figure 5 of \cite{li:2015}, the propyne/propadiene ratio is most sensitive to the hydrogen abundance above 500 km - hence we are not able to precisely predict the atomic hydrogen abundance from these observations.  To more accurately predict the concentration of H, we require measurements at higher altitudes - either through higher spatial resolution or higher spectral resolution studies.
	
	\section{Conclusion}
	The detection and measurement we present in this Letter serve to provide a constraint on current photochemical models of Titan's atmosphere.  We recommend that models take into careful consideration the isomeric structure of molecules, as each structure may be subject to different possible production and destruction pathways.  An important goal for future studies will be to extend this work to measure the abundance of propadiene in polar regions on Titan, where unique chemical concentrations and reactions occur.

	\section{Acknowledgments}
	N.A.L., C.A.N., and F.M.F. were supported by the NASA Cassini Project for the research work reported in this paper.  P.G.J.I. and N.A.T were funded by the UK Science and Technology Facilities Council.  N.A.L. was supported by NASA through the CRESST II cooperative agreement CA 80GSFC17M0002.  C.A.N. received additional funding through NASA's Astrobiology Institute grant "Habitability of Hydrocarbons Worlds: Titan and Beyond", and NASA's Solar System Observations Program.  N.A.L. and T.K.G were Visiting Astronomers at the Infrared Telescope Facility, which is operated by the University of Hawaii under contract NNH14CK55B with the National Aeronautics and Space Administration.

	\bibliography{allene}

\end{document}